\documentclass[journal abbreviation]{copernicus}

\begin{document}
\begin{nolinenumbers} % arXiv hates line numbers

\title{Efficiency and robustness in Monte Carlo sampling
       of 3-D geophysical inversions with Obsidian v0.1.2:
       Setting up for success}

% \Author[affil]{given_name}{surname}

\Author[1]{Richard}{Scalzo}
\Author[2]{David}{Kohn}
\Author[3]{Hugo}{Olierook}
\Author[4]{Gregory}{Houseman}
\Author[1,5]{Rohitash}{Chandra}
\Author[6,7]{Mark}{Girolami}
\Author[1,8]{Sally}{Cripps}

%% The [] brackets identify the author with the corresponding affiliation.
%% 1, 2, 3, etc. should be inserted.

\affil[1]{Centre for Translational Data Science, University of Sydney,
          Darlington NSW 2008, Australia}
\affil[2]{Sydney Informatics Hub, University of Sydney,
          Darlington NSW 2008, Australia}
\affil[3]{School of Earth and Planetary Sciences, Curtin University,
          Bentley WA 6102, Australia}
\affil[4]{School of Earth and Environment, University of Leeds,
          Leeds, LS2 9JT, UK}
\affil[5]{School of Geosciences, University of Sydney,
          Darlington NSW 2008, Australia}
\affil[6]{The Alan Turing Institute for Data Science,
          British Library, 96 Euston Road, London, NW1 2DB, UK}
\affil[7]{Department of Mathematics,
          Imperial College London, London, SW7 2AZ, UK}
\affil[8]{School of Mathematics and Statistics,
          University of Sydney, Darlington NSW 2008, Australia}

\runningtitle{3-D geophysical inversions with \emph{Obsidian} v0.1.2}
\runningauthor{Scalzo et al.}
\correspondence{Richard Scalzo (richard.scalzo@sydney.edu.au)}

\received{}
\pubdiscuss{} %% only important for two-stage journals
\revised{}
\accepted{}
\published{}

%% These dates will be inserted by Copernicus Publications during the typesetting process.

\firstpage{1}

\maketitle

\begin{abstract}
The rigorous quantification of uncertainty in geophysical inversions is a
challenging problem. Inversions are often ill-posed and the likelihood surface
may be multimodal; properties of any single mode become inadequate uncertainty
measures, and sampling methods become inefficient for irregular posteriors
or high-dimensional parameter spaces.  We explore the
influences of different choices made by the practitioner on the efficiency and
accuracy of Bayesian geophysical inversion methods that rely on Markov chain
Monte Carlo sampling to assess uncertainty, using a multi-sensor inversion
of the three-dimensional structure and composition of a region in the
Cooper Basin of South Australia as a case study.  The inversion is performed
using an updated version of the Obsidian distributed inversion software.
We find that the posterior for this
inversion has complex local covariance structure, hindering the efficiency of
adaptive sampling methods that adjust the proposal based on the chain history.
Within the context of a parallel-tempered Markov chain Monte Carlo scheme for
exploring high-dimensional multi-modal posteriors, a preconditioned
Crank-Nicholson proposal outperforms more conventional forms of random walk.
Aspects of the problem setup, such as priors on petrophysics or on 3-D
geological structure, affect the shape and separation of posterior modes,
influencing sampling performance as well as the inversion results.  Use of
uninformative priors on sensor noise can improve inversion results by enabling
optimal weighting among multiple sensors even if noise levels are uncertain.
Efficiency could be further increased by using posterior gradient information
within proposals, which Obsidian does not currently support, but which could
be emulated using posterior surrogates.
\end{abstract}

\copyrightstatement{TEXT}

% ============================================================================

\introduction  %% \introduction[modified heading if necessary]

Construction of 3-D geological models is plagued by the limitations
{on direct sampling and geophysical measurement}
\citep{wellmann2010,lindsay2013}.
% \citep{pan2012,wang2012}.
{Direct geological observations are sparse because of the difficulty
in acquiring them, being often obscured by sedimentary or regolith
cover; resolving this issue via drilling is expensive
\citep{anand2010,salama2016}.}
Indirect observations via geophysical sensors deployed at or above the surface
are more readily obtained
\citep{strangway1973,gupta1985,sabins1999,nabighian2005mag,nabighian2005grav}.
However, {gravity, magnetic, and electrical measurements} integrate
data from the surrounding volume, so it is difficult to resolve precise
geological constraints at any given position and depth,
{except where borehole measurements are also available.}
Determining the true underlying geological structure, or range of geological
structures, consistent with observations constitutes
an {often poorly constrained} inverse problem.
One natural way to approach this is forward-modelling, where the responses of
various sensors on a proposed geological structure are simulated, and the
proposed structure is then updated or sampled iteratively
\citep[for examples see][]{jessell2001,calcagno2008,olierook2015}.

The incompleteness and uncertainty of the information contained in any
single data set frequently mean that there are many possible worlds consistent
with the data being analyzed
\citep{tarantola1982generalized,tarantola2005inverse}.
To the extent that information provided by different datasets is
complementary, combining {all available} information into
a single joint inversion reduces uncertainty in the final results.
Accomplishing this in a principled and
self-consistent manner presents several challenges, including:
(i) how to weigh constraints provided by different datasets
relative to each other;
(ii) how to rule out worlds inconsistent with geological processes
(expert knowledge);
(iii) how to present a transparent accounting of the remaining uncertainty;
and (iv) how to do all this in a computationally efficient manner.

Bayesian statistical techniques provide a powerful framework for
characterizing and fusing disparate sources of probabilistic information
\citep{tarantola1982generalized,mosegaard1995monte,
       sambridge2002monte,sambridge2012}.
All input sources of information --- from geophysical sensors, geological
field observations, previous inferences, or expert knowledge --- are treated
as probability distributions; this forces the practitioner
to make explicit all assumptions, not only about expected values, but about
uncertainties.  The output of a Bayesian method is also a probability
distribution (the \emph{posterior}),
{for which the gold-standard representation is}
a set of samples from a Monte Carlo algorithm, in particular
\emph{Markov chain Monte Carlo} \citep[MCMC;][]
    {mosegaard1995monte,sambridge2002monte}.
The posterior distribution is a representation of all possible outcomes and
hence provides an internal estimate of uncertainty.
The uncertainty associated with the posterior can be visualized
{in terms of the marginal distributions of parameters of interest,}
or rendered in 3-D voxelisations of information entropy
\citep{wellmann2012uncertainties}.  The posterior also can be readily updated
online as new information becomes available, making Bayesian approaches
optimal for decision-making under risk and uncertainty.

Although Bayesian methods provide rigorous uncertainty quantification,
implementing them in practice for complicated forward models with many
free parameters has proven difficult in other geoscientific contexts, such as
landscape evolution \citep{chandra2018bayeslands} and coral reef assembly
\citep{JPall_BayesReef2018}.  \citet{sambridge2002monte}
point out the challenge of capturing all elements of
a geophysical problem in terms of probability, which can be difficult for
complex datasets and even harder for approximate forward models or world
representations where the precise nature of the approximation is hard to
capture.  The irregular shapes and multimodal structure of the posterior
distributions for realistic geophysics problems makes them hard to explore;
the second moment (variance) of the posterior around each
{local maximum} may in these cases significantly underestimate
uncertainties.  Moreover, the large number of parameters
{needed to specify} 3-D structures also means these irregular
posteriors are embedded in high-dimensional spaces, increasing
the computational cost for both optimization and sampling.
Therefore, the sampling methods must usually be tailored to each
individual problem and no {``one-size-fits-all'' solution exists.}

These limitations form the backdrop for current work on applying Bayesian
principles to 3-D structural modelling.  \citet{giraud2017,giraud2018}
demonstrate an optimization-based Bayesian inversion framework
{for 3-D geological models}, which finds
the maximum of the posterior distribution
(\emph{maximum a posteriori}, or MAP), and expresses uncertainty in terms of
the posterior covariance around the MAP solution; while they show that
fusing data reduces uncertainty around this mode, they do not attempt to
find or characterize other modes, or higher moments of the posterior.
\citet{ruggeri2015} investigate several MCMC schemes for sampling a
single-sensor inverse problem
{(crosshole georadar travel time tomography)},
focusing on sequential, localized perturbations of a proposed 3-D
model (``sequential geostatistical resampling'', or SGR); they show that
sampling is impractically slow due to high dimensionality and correlations
between model parameters.  \citet{laloy2016} embed the SGR proposal within a
parallel-tempered sampling scheme to explore multiple posterior modes
{of a 2-D inverse problem in groundwater flow}, improving
computational performance but not to a {cost-effective} threshold.
The above methods are non-parametric, in that the model parameters simply
form a 3-D field of rock properties to which sensors respond.
\citet{delavarga2016,gempy} focus on building parametrized 3-D models
in order to reduce the problem dimension and to naturally incorporate
structural measurements, but have not yet tested this framework on a
large-scale 3-D joint inversion with multiple sensors.

\citet{mccalman2014} present Obsidian, a flexible software platform for
MCMC sampling of 3-D multi-modal geophysical models on distributed computing
clusters.  \citet{beardsmore2016} demonstrate Obsidian on a test problem in
geothermal exploration, in the Moomba gas field of the Cooper Basin
in South Australia, comparing {their results}
to a deterministic inversion of the same area
performed by \citet{meixner2009}.  These papers outline a full-featured
open-source inversion method that can fuse heterogeneous data into a
detailed solution, but make few comments about how the efficiency and
robustness of the method depends on the particular choices they made.

In this paper, we revisit the inversion problem of \citet{beardsmore2016}
using a customized version of the \citet{mccalman2014} inversion code.
Our interest is in exploring this problem as a case study to determine
which aspects of this problem's posterior present the most significant
obstacles to efficient sampling, which updates to the MCMC scheme improve
sampling under these conditions, and how plausible alternative choices of
problem setup might influence the efficiency of sampling or the robustness
of the inversion.  The aspects we consider include:  correlations
between model parameters; relative weights between datasets with poorly
constrained uncertainty; and choices of priors representing different
possible exploration scenarios.

% ============================================================================

\section{Background}
\label{sec:background}

In this section we present a brief overview of the Bayesian forward-modeling
paradigm to geophysical inversions. We also provide a discussion of
implementing Bayesian inference via sampling using MCMC methods.
We then present background on the original Moomba
inversion problem, commenting on choices made in the {inversion process}
before we begin to explore different choices in subsequent sections.

% ----------------------------------------------------------------------------

\subsection{Overview of Bayesian inversion}
\label{subsec:bayes}

A Bayesian inversion scheme for geophysical forward models comprises of
three key elements:
\begin{enumerate}
\item the underlying parametrized representation of the simulated volume
      or history, which we call the \emph{world} or \emph{world view},
      denoted by {a vector of} \emph{world parameters}
      $\vec{\theta} = (\theta_1, \ldots, \theta_P)$
\item a probability distribution $p(\vec{\theta})$ over the world parameters,
      called the \emph{prior}, expressing expert knowledge or belief
      about the world before any datasets are analyzed; and
\item a probability distribution $p(\mathcal{D}|\vec{\theta})$ over
      possible realizations of the observed data $\mathcal{D}$ as a function
      of world parameters, called the \emph{likelihood}, that incorporates
      the prediction of a deterministic forward model $g(\vec{\theta})$
      of the sensing process for each value of $\vec{\theta}$.
\end{enumerate}
The \emph{posterior} is then the distribution $p(\vec{\theta}|\mathcal{D})$
of values of the world parameters consistent with both prior knowledge and
observed data.  \emph{Bayes' theorem} describes the relationship between
the prior, likelihood, and posterior:
\begin{equation}
p(\vec{\theta}|\mathcal{D})
   = \frac{p(\mathcal{D}|\vec{\theta})p(\vec{\theta})}
          {\int p(\mathcal{D}|\vec{\theta})p(\vec{\theta})\,d\vec{\theta}}.
\label{eqn:bayes}
\end{equation}

{Our terminology for these elements is typical of the statistics
literature, so it is critical to identify the same elements in terminology
used in previous geophysical inversion papers
\citep[for example][]{menke2018}.  In previous papers a ``model''
might refer to the world representation, whereas below we will use the word
``model'' to refer to the \emph{statistical} model defined by a choice of
all of the above elements.
A non-Bayesian inversion would proceed by minimizing an
\emph{objective function}, one simple form of which is the mean square misfit
between the (statistical) model predictions and the data, corresponding to
our negative log likelihood (for observational errors that are independent and
Gaussian-distributed with precisely known variance).  To penalize solutions
that are considered \emph{a priori} unlikely,
the objective function might include additional
\emph{regularization} terms corresponding to the negative log priors in our
framework.  The full objective function would thus correspond to our negative
log posterior, and minimization of the objective function would correspond to
maximization of our posterior probability, under some choice of prior.
However, regularization does not necessarily proceed from a probabilistic
interpretation; interpreting model elements in terms of probability may
motivate different choices of likelihood or prior than the
usual non-probabilistic misfit or regularization terms.}

Indeed, there is considerable flexibility in choosing the above elements
even in a fully probabilistic context.
For example, the partitioning of information into ``data'' and
``prior knowledge'' is neither unique nor cut-and-dried.  However,
there are guiding principles:  the ideal set of parameters $\vec{\theta}$ is
both \emph{parsimonious} --- as few as possible to faithfully represent
the world --- and \emph{interpretable}, referring to meaningful aspects
of the world that can easily be read off the parameter vector.
Information resulting from processes that can be easily simulated belong
in the likelihood:  for example, one might argue that the output of a
{gravimeter} should have a Gaussian distribution,
because it responds to the
mean rock density within a volume and hence obeys the central limit theorem,
or that the output of a Geiger counter should follow a Poisson distribution
to reflect the physics of radioactive decay.  Even processes that are not so
easily simulated can at least be approximately described, for example by
using a mixture distribution {to account for outlier measurements}
\citep{mosegaard1995monte}
or a prior on the unknown noise level in a process \citet{sambridge2012}.
Other information about allowable or likely worlds
belongs in the prior, such as the distribution of initial
conditions for simulation, or interpretations of datasets with expensive or
intractable forward models.

The inference process expresses its results in terms either of
$p(\vec{\theta}|\mathcal{D})$ itself or of integrals over
$p(\vec{\theta}|\mathcal{D})$ (including credible limits on $\vec{\theta}$).
This is different from the use of point estimates for the world parameters,
such as the \emph{maximum likelihood} (ML) solution
$\vec{\theta}_\mathrm{ML} = \sup_{\vec{\theta}} p(\mathcal{D}|\vec{\theta})$
or the \emph{maximum a posteriori} (MAP) solution
$\vec{\theta}_\mathrm{MAP} =
 \sup_{\vec{\theta}} p(\mathcal{D}|\vec{\theta}) p(\vec{\theta})$.
To the extent that ML or MAP prescriptions give any estimate of uncertainty
on $\vec{\theta}$, they usually do so through {the covariance}
of the log likelihood or {log}
posterior around the optimal value of $\vec{\theta}$, equivalent to a local
approximation of the likelihood or posterior by a multivariate Gaussian.
As mentioned above, these approaches will underestimate the uncertainty for
complex posteriors; a more rigorous accounting of uncertainty will include all
known modes, higher moments of the distribution, or (more simply) providing
enough samples from the distribution to characterize it.

{The posterior distribution $p(\vec{\theta}|\mathcal{D})$
is rarely available in closed form.  However, it is often known up to
a normalizing constant:  $p(\mathcal{D}|\vec{\theta}) p(\vec{\theta})$.
Sampling methods such as MCMC can therefore be used to approximate the
posterior, without having to explicitly evaluate the normalizing constant
(the high-dimensional integral in the denominator
of Eq.~\ref{eqn:bayes}).  It is to these methods we turn next.}

% ----------------------------------------------------------------------------

\subsection{Markov chain Monte Carlo}
\label{subsec:mcmc}

A MCMC algorithm comprises a sequence of world parameter vectors
$\{ \vec{\theta}^{[j]} \}$, called a \emph{(Markov) chain},
and a \emph{proposal distribution} $q(\vec{\theta}'|\vec{\theta})$ to generate
a new set of parameters based only on the last element of the chain.
In the commonly-used \emph{Metropolis-Hastings algorithm}
\citep{metropolis1953equation,hastings1970monte}, a proposal
$\vec{\theta}' \sim q(\vec{\theta}'|\vec{\theta}^{[j]})$
is at random either \emph{accepted} and added to the chain's history
($\vec{\theta}^{[j+1]} = \vec{\theta}'$) with probability
\begin{equation}
P_\mathrm{accept} = {\min} \left( 1,
    \frac{P(\mathcal{D}|\vec{\theta}') P(\vec{\theta}')
          q(\vec{\theta}^{[j]} |\vec{\vec{\theta}}')}
         {P(\mathcal{D}|\vec{\theta}^{[j]}) P(\vec{\theta}^{[j]})
          q(\vec{\theta}'|\vec{\theta}^{[j]})} \right),
\end{equation}
or \emph{rejected} and a copy of the previous state added instead
($\vec{\theta}^{[j+1]} = \vec{\theta}^{[j]}$).
This rule guarantees, under certain regularity
conditions \citep{chibgreenberg95}, % ask Sally about this
that the sequence $\{ \vec{\theta}^{[j]} \}$
converges to the required stationary distribution,
$P(\vec{\theta}|\mathcal{D})$, in the limit of increasing $n$.

Metropolis-Hastings algorithms form a large class of sampling algorithms,
limited only by the forms of proposals.
Although proofs that the chain will \emph{eventually} sample from the
posterior are important, clearly chains based on \emph{efficient} proposals
are to be preferred.  A proposal's efficiency will depend on
the degree of correlation between consecutive states in the chain, which in
turn can depend on how well matched the proposal distribution is to the
properties of the posterior.

One simple, commonly used proposal distribution is a (multivariate)
\emph{Gaussian random walk} (GRW) step $u$ from the chain's current position,
{drawn from a multivariate Gaussian distribution}
with covariance matrix $\vec{\Sigma}$:
\begin{equation}
\vec{\theta}' = \vec{\theta}^{[j]} + \vec{u},
    \hspace{0.5in} \vec{u} \sim N(\vec{0}, \vec{\Sigma}).
\end{equation}
This proposal is straightforward to implement, but its effectiveness can
depend strongly on $\vec{\Sigma}$, and does not in general scale well to rich,
high-dimensional world parametrizations.  If $\vec{\Sigma}$ has too large a
scale, the GRW proposal will step too often into regions of low probability,
resulting in many repeated states due to rejections; if the scale is
too small, the chain will take only small, incremental steps.
In both cases, subsequent states are highly correlated.
If the shape of $\vec{\Sigma}$ is not tuned to capture
correlations between different dimensions of $\vec{\theta}$, the overall
scale must usually be reduced to ensure a reasonable acceptance fraction.

The SGR method \citep{ruggeri2015,laloy2016} can be seen as a mixture of
multivariate Gaussians, in which $\vec{\Sigma}$ has highly correlated
sub-blocks of parameters, corresponding to {variations of the world
over different spatial scales.}
{\citet{ruggeri2015} and \citet{laloy2016} evaluate SGR using
single-sensor inversions in crosshole georadar travel time tomography,
with posteriors corresponding} to a Gaussian process ---
an unusually tractable (if high-dimensional) problem that could be solved
in closed form as a cross-check.  {These authors} found that in general
{updating blocks of parameters simultaneously}
was inefficient, which may not be surprising in a high-dimensional model:
for a tightly constrained posterior lying along a low-dimensional subspace
of parameter space, almost all directions --- hence almost all posterior
covariance choices --- lead towards regions of low probability.  Directions
picked at random without regard for the shape of the posterior will scale
badly with increasing dimension.

Many other types of proposals exist, using information from ensembles of
particles \citep{goodmanweare2010affine}, adaptation of the proposal
distribution based on the chain's history \citep{haario2001adaptive},
derivatives of the posterior \citep{neal2011mcmc,girolami2011riemann},
approximations to the posterior \citep{strathmann2015kmc}, and so forth.
The GRW proposal is not only easy to write down and fast to evaluate,
but requires no derivative information.  We will compare and contrast several
derivative-free proposals in our experiments below.

The posterior distributions arising in geophysical inversion problems are
also frequently multi-modal; MCMC algorithms to sample such posteriors
need the ability to escape from, or travel easily between, local modes.
\emph{Parallel-tempered MCMC} PTMCMC \citep{Geyer95} is a meta-method for
sampling multi-modal distributions that works by running an ensemble of
Markov chains.  The ensemble is characterized by a sequence of
{$M+1$} parameters
$\{ \beta_i \}$, with $\beta_0 = 1 > \beta_1 > \beta_2 > \ldots > \beta_M >
0$,
called the \emph{(inverse) temperature ladder}.
Each chain samples the distribution
\begin{equation}
P_i(\vec{\theta}|D) \propto (P(D|\vec{\theta}))^{\beta_i} P(\vec{\theta}),
\end{equation}
so that the chain with $\beta_0 = 1$ is sampling from the desired posterior,
and a chain with $\beta_i = 0$ samples from the prior, which should be easy
to explore.  Chains with intermediate values $0 < \beta < 1$ sample
intermediate distributions in which the data's influence is reduced,
so that modes are shallower and easier for chains to escape and traverse.
In addition to proposing new states within each chain, PTMCMC includes
{Metropolis-style} proposals that allow adjacent chains on the
temperature ladder, {with inverse temperatures $\beta$ and $\beta'$,
to swap their most recent states $\vec{\theta}$ and $\vec{\theta}'$
with probability
\begin{equation}
P_{\mathrm{swap}} = \min \left( 1,
    \left[ \frac{P(\mathcal{D}|\vec{\theta}')}
                {P(\mathcal{D}|\vec{\theta})} \right]^{\beta' - \beta}
    \frac{P(\vec{\theta}')}{P(\vec{\theta})}
\right).
\end{equation}
This allows chains with current states spread throughout parameter space
to share global information about the posterior} in such a way that
chain $i$ still samples $P_i(\vec{\theta}|D)$ in the long-term limit.
{The locations of discovered modes diffuse from low-$\beta_i$ chains
(which can jump freely between relaxed, broadened versions of these modes)
towards the $\beta_0 = 1$ chain, which can then sample from all modes of the
unmodified posterior in the correct proportions.}
The temperature ladder should be defined so that adjacent chains on the ladder
are sampling from distributions similar enough for swaps to occur frequently.

Figure~\ref{fig:ptmodes}
{illustrates the sampling of a simple bimodal probability distribution
      (a mixture of two Gaussians) via PTMCMC}.
The solid line depicts the true bimodal distribution, while the broken lines
shows the stationary distribution of tempered chains for smaller values
of $\beta$.  The tempered chains are more likely to propose moves across
modes than the untempered chains, and the existence of a sequence of chains
ensures that the difference in probability between successive chains is small
enough that swaps can take place easily.

% Width below is either 8.3 cm (one-column) or 12 cm (two-column).
\begin{figure}[t]
\includegraphics[width=8.3cm]{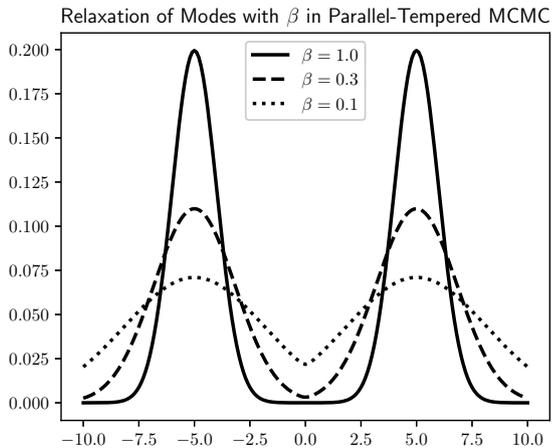}
\caption{Parallel-tempered relaxation of a bimodal distribution.}
\label{fig:ptmodes}
\end{figure}

Even without regard to multiple modes, PTMCMC can also help to reduce
{correlations} between successive independent posterior samples.
\citet{laloy2016} use SGR as a within-chain proposal in a PTMCMC scheme,
demonstrating its effects on {correlations} between samples but
noting that the algorithm remains computationally intensive.

% ----------------------------------------------------------------------------

\subsection{Performance metrics for MCMC}
\label{subsec:metrics}

Because MCMC guarantees results only in the limit of large samples,
criteria are still required to assess the algorithm's performance.
{Suppose for the discussion below that up to the assessment point,
we have obtained $N$ samples of a $d$-dimensional posterior from each of
$M$ separate chains; let
$\vec{\theta}^{[j]}_i = (\theta^{[j]}_{1i}, \ldots, \theta^{[j]}_{di})$
be the $d \times 1$ vector of parameter values drawn at iteration $[j]$
in chain $i$. Let 
\[
\hat{\theta}_{ki}=\frac{1}{N} \sum_{j=1}^N \theta^{[j]}_{ki}
\] 
be the mean value of the parameter $\theta_k$ in chain $i$, across the $N$
iterates, and let $\tilde{\theta}_k=\frac{1}{M}\sum_{i=1}^M\hat{\theta}_{ki}$
be the sample mean of $\theta_k$ across all iterates and chains. Then
\[
B_{k}=\frac{1}{M-1}\sum_{i=1}^M(\hat{\theta}_{ki}-\tilde{\theta}_{k})^2
\]
Further define
\[
s^2_{ki}=\frac{1}{M-1}\sum_{j=1}^N(\theta^{[j]}_{ki}-\hat{\theta}_{ki})^2
\] and
\[
W_{k}=
\frac{1}{M}\sum_{i=1}^M s^2_{ki}.
\]
}

For Metropolis-Hastings MCMC, the \emph{acceptance fraction} of proposals is
easily measured, and for a chain that is performing well should be
$\sim 20$--$50\%$.  {\citet{roberts1997weak} showed that the optimal
acceptance fraction for random walks in the limit of a large number of
dimensions} is 0.234, which we will take as our target since
the proposals we will consider are modified random walks.

{We examine correlations between samples within each chain separated
by a lag time $l$ using the \emph{autocorrelation function},
\begin{equation}
\rho_{lki} = \frac{1}{(N-l) W_k}
\sum_{j=l+1}^N (\theta^{[j]}_{ki} - \hat{\theta}_{ki})
               (\theta^{[j-l]}_{ki} - \hat{\theta}_{ki}),
\end{equation}
The number of independent draws from the posterior with equal statistical
power to each set of $N$ chain samples scales with the area under the
autocorrelation function or \emph{(integrated) autocorrelation time} (IACT),
\begin{equation}
\tau_{ki} = 1 + 2 \sum_{l=1}^N \left( 1 - \frac{k}{N} \right) \rho_{lki}.
\end{equation}
A \emph{trace plot} of the history of an element of the parameter vector
$\vec{\theta}$ over time summarizes the sampling performance at a glance,
revealing where in parameter space an algorithm is spending its time;
Fig.~\ref{fig:traceplot} shows a series of such figures for some
of the different MCMC runs in the present work.}

{\citet{gelmanrubin1992} assess the number of samples required to reach
a robust sampling of the posterior by comparing results among multiple chains.
If the simulation has run long enough, the mean values among chains
should differ by some small fraction of the width of the distribution;
intuitively, this is similar to a hypothesis test that the chains are
sampling the same marginal distribution for each parameter.
More precisely, the quantity
\begin{equation}
\hat{V}_k/W_k = \frac{N-1}{N} + \frac{M + 1}{MN} B_k/W_k
\label{eqn:gelmanrubin}
\end{equation}
provides a metric for convergence of different chains to the same
result, which decreases to 1 as $N \rightarrow \infty$.}
The chains may be stopped and results read out when the metric dips below
a target value for all world parameters $\vec{\theta}$.
{The precise number of samples needed may depend on the details of
the distribution; the metric provides a stopping condition, but not an
estimate of how long it will take to achieve.}

The results from this procedure must still be evaluated according to how well
the underlying statistical model describes the geophysical data, and whether
the results are geologically plausible --- although this is not unique to MCMC
solutions.  The distribution of residuals of model predictions
(forward-modeled data sets) from the observed data can be compared to the
assumed likelihood.  The standard deviation or variance of the residuals
{(relative to the uncertainty)} provide a convenient single-number summary,
but the spatial distribution of residuals may also be important; outliers
and/or structured residuals will indicate places where the model fails
to predict the data well, and highlight
{parts of the model parametrization that need refinement.}

Finally, {representative instances of}
the world itself should be visualized to check for surprising features.
Given the complexity of real-world data, the adequacy of a given model is in
part a matter of scientific judgment, or fitness for a particular applied
purpose to which the model will be put.
{We will use the term \emph{model inadequacy} to refer to model errors
arising from approximations or inaccuracies in the world parametrization
or the mathematical specification of the forward model --- although there will
always be such approximations in real problems, and the presence of model
inadequacy should not imply that the model is unfit for purpose.}

% ----------------------------------------------------------------------------

\subsection{The Obsidian distributed PTMCMC code}
\label{subsec:obsidian}

For our experiments we use a customized fork {(v0.1.2)} of the
open-source Obsidian software package.
{Obsidian was previously presented in \citet{mccalman2014} and was
used to obtain the modeling results of \citet{beardsmore2016}; v0.1.1 was
the most recent open-source version publicly available before our work.}
We refer the reader to {previous publications} for a comprehensive
description of Obsidian, but below we summarize key elements corresponding
to the inversion framework set out above.

{\bf World parametrization:} 
Obsidian's world is parametrized as a series of discrete layers, each with
its own spatially constant rock properties, separated by smooth boundaries.
Each layer boundary is a two-dimensional Gaussian process regression against
a set of \emph{control points} that specify the subsurface depth of the
boundary at given surface locations.  The layer boundaries are indexed in
order of increasing depth beneath the surface, but are allowed to cross
over each other.  In regions where the formal layer thickness $z_{i+1} - z_i$
is negative, the corresponding rock layer pinches out to zero thickness.
For a world with $N$ layers, indexed by $i$ with $1 \leq i \leq N$,
each with {a grid $n_i$ regularly spaced} control points at sites $x_i$
and rock properties corresponding to each of $S$ forward-modeled sensors,
the parameter vector is therefore
\begin{equation}
\vec{\theta} = ( \alpha_{11} \ldots \alpha_{Nn_N},
                 \rho_{11} \ldots \rho_{NK} ),
\end{equation}
where $\alpha_{ij}$ is the offset of the mean depth of the top of layer $i$
at site $j$, and $\rho_{is}$ is the rock property of layer $i$ associated
with sensor $s$.  Taken together, the rock properties for each layer and the
control points for the boundaries between the layers fully specify the world.
{This parametrization requires that interface depths be single-valued,
not for example permitting the surface to fold above or below.  Such a
limitation still enables reasonable representations of sedimentary basins,
but may hinder faithful modeling of other kinds of structures.}

{\bf Prior:}
The control point depth offsets within each layer $i$ have a multivariate
Gaussian prior with mean zero and covariance $\vec{\Sigma_{\alpha_i}}$.
The Gaussian processes which interpolate the layer boundaries across the
lateral extent of the {world} use a radial basis function kernel,
\begin{equation}
k(x, y; x', y') = \exp \left(
  -\frac{(x-x')^2}{\Delta_x^2}
  -\frac{(y-y')^2}{\Delta_y^2} \right),
\end{equation}
and has mean function $\mu_i(x, y)$ that can be specified at finer resolution
to capture fine detail in layer structure.
The correlation lengths $\Delta_x$ and $\Delta_y$ could in principle
be varied, but in this case are fixed in value to the spacing between
control point locations along the $x$ and $y$ coordinate axes, respectively.
The rock properties for each layer $i$, which are statistically independent
of the control points, also have a multivariate Gaussian prior, with mean
$\vec{\mu}_{\rho_i}$ and covariance $\vec{\Sigma}_{\vec{\rho}_i})$.
The prior for the full parameter vector is therefore block-diagonal,
\begin{eqnarray}
P(\vec{\theta}) & = & \prod_{i=1}^N
                P(\vec{\alpha}_{i\cdot})
                P(\vec{\rho}_{i\cdot}) \nonumber \\
          & = & \prod_{i=1}^N
                N(\vec{\alpha}_{i\cdot}; 0, \vec{\Sigma}_{\vec{\alpha}_i})
                N(\vec{\rho}_{i\cdot}; \vec{\mu}_{\vec{\rho}_i},
                                       \vec{\Sigma}_{\vec{\rho}_i}).
\end{eqnarray}

{\bf Likelihood:}
The likelihood for each Obsidian sensor $s$ is Gaussian, meaning that
the residuals of the data $\mathcal{D}_s$ from the forward model predictions
$f_s(\vec{\theta})$ for the true world parameters $\theta$ are assumed to be
independent, identically distributed Gaussian draws.
The underlying variance of the Gaussian noise is not known, but is assumed to
follow an inverse gamma distribution $\mathrm{IG}(x;\alpha_s,\beta_s)$
with different (user-specified) hyperparameters $\alpha_s$, $\beta_s$
for each sensor $s$.
This choice of distribution amounts to a prior, but the hyperparameters
$\alpha_s$ and $\beta_s$ for each sensor are not explicitly sampled over;
instead, they are integrated out analytically, so that the final likelihood
has the form
\begin{equation}
P(\mathcal{D}_s|\vec{\theta}) = \prod_{k=1}^{K_s} t_{2\alpha_s} \left(
    \frac{\beta_s}{\alpha_s} (f_s(\vec{\theta}) - \mathcal{D}_s) \right),
\end{equation}
where $t_\nu (x)$ is a Student's-$t$ distribution with $\nu$
degrees of freedom.  This distribution is straightforward to calculate,
although the results may be sensitive to the user's choices of $\alpha_s$
and $\beta_s$; unrestrictive choices (e.g. $\alpha_s = \beta_s = 1$)
should be used if the user has little prior knowledge about the noise level
in the data.
The likelihood including all sensors is therefore
\begin{equation}
P(\mathcal{D}|\vec{\theta}) = \prod_{s=1}^S P(\mathcal{D}_s|\vec{\theta}),
\end{equation}
since each sensor probes a different physical aspect of the rock.

{\bf MCMC:} The sampling algorithm used by Obsidian is an adaptive form of
PTMCMC, described in detail in \citet{ptadapt}.
This algorithm allows for the progressive adjustment of the step size used for
proposals within each chain, as well as the temperature ladder used to sample
across chains, as sampling progresses.  A key feature of the adjustment
process is that the maximum allowed change to any chain property diminishes
over time, made inversely proportional to the number of samples;
this is necessary to ensure that the chains converge to the correct
distribution in the limit of
large numbers of samples \citep{robertsrosenthal2007adaptive}.
The Obsidian implementation of PTMCMC also allows it to be run on
distributed computing clusters, making it truly parallel in resource use
as well as in the requirement for multiple chains.

% ----------------------------------------------------------------------------

\subsection{The original Moomba inversion problem}
\label{subsec:moomba}

The goal of the original Moomba inversion problem
\citep{beardsmore2016,mccalman2014} was to identify potential geothermal
energy applications from hot granites in the South Australian part of the
Cooper Basin (cf. \citet{carr2016onshore} for a recent review of the
Cooper Basin). Modeling the structure of granite intrusions and their
temperature enabled the inference of the probability of the presence of
granite above 270~$^{\degree}$C at any point within the volume.
The chosen region was a portion of the Moomba gas field with dimensions of
35~$\times$~35~$\times$~12~km volume centered at
~-28.1$^{\degree}$~S, ~140.2$^{\degree}$~E.
The volume is divided into six layers, with the first four being thin,
{sub-}horizontal, Permo--Triassic sedimentary layers, the fifth
corresponding to Carboniferous--Permian granitoid intrusions (Big Lake Suite),
and the sixth to a Proterozoic basement \citep{carr2016onshore}.
The number of layers and the priors on mean depths of layer boundaries were
related to interpretations of depth-converted seismic reflection horizons
published by the Department of State Development (DSD) in South Australia
\citep{beardsmore2016}.
Data used in the inversion include {Bouguer} anomaly;
{total magnetic intensity};
magnetotelluric sensor data; temperature measurements from gas wells;
and petrophysical laboratory measurements based on 115 core samples
from holes drilled throughout the region.  Rock properties measured for each
sample include density, magnetic susceptibility, thermal conductivity,
thermal productivity, and resistivity.

The original choices of how to partition knowledge between prior and
likelihood struck a balance between
{accuracy of the world representation} and computational efficiency.
The empirical covariances of the petrophysical sample measurements for each
layer were used to specify a multivariate Gaussian prior on that layer's
rock properties; although these measurements could be construed as data,
the simplifying assumption of spatially constant mean rock properties
left little reason to write their properties into the likelihood.
The gravity, magnetic, magnetotelluric, and thermal data all directly
constrained rock properties relevant to the geothermal application and were
explicitly forward-modeled as data. ``Contact points'' from drilled wells,
directly constraining the layer depths in the neighborhood of a drilled hole
as part of the likelihood, were available {and used to inform the prior,
but not treated as sensors in the likelihood.}
Treating the seismic measurements as data would have dramatically increased
computational overhead relative to the use of interpreted reflection horizons
as mean functions for layer boundary depths in the prior.  Using interpreted
seismic data to inform the mean functions of the layer boundary priors also
reduced the dimension of the parameter space, letting the control points
specify long-wavelength deviations from {seismically derived}
prior knowledge: each reflection horizon was interpolated onto
a $20 \times 20$ grid, meaning that 400 control points per layer
(resulting in 2400 parameters for the world geometry alone) would have been
required to define the high-resolution reference world.

Given this knowledge of the local geology
\citep{carr2016onshore,mccalman2014distributed},
the world parameters for geometry were chosen as follows:
The surface was fixed by a level plane at zero depth.
The control point grids for the relatively simple sedimentary layers were
specified by $2 \times 2$ grids of control points (lateral spacing: 17.5~km).
The layer boundary for the granite intrusion layer used a $7 \times 7$ grid
(lateral spacing:  5~km), and also underwent a nonlinear transformation
stretching the boundary vertically, to better represent the elongated shapes
of the intrusions.  Including the rock properties, this allowed the entire
world to be specified by a vector of 101 parameters, a large but not
unmanageable number.

Figure~\ref{fig:moomba-slices} show horizontal slices through the posterior
probability density for granite at a depth of 3.5 km, similar to that shown
in figure~9 of \citet{beardsmore2016}, for three MCMC runs sampling from the
original problem.  While the posterior samples from the previous inference
are not available for quantitative comparison, we see reasonable qualitative
agreement with previous results in the cross-sectional shape of the granite
intrusion.

% ============================================================================

\section{Experiments}

To demonstrate the impact of problem setup and proposal efficiency in a
Bayesian MCMC scheme for geophysical inversion, we run a series of experiments
altering the prior, likelihood, and proposal for the Moomba problem.
We approach this variation as an iterative investigation into the nature of
the data and the posterior's dependence on them, motivating each choice with
the intent of relating our findings to related 3-D inversion problems.

The datasets we use for our experiments are the gravity anomaly, total
magnetic intensity, and magnetotelluric readings originally distributed as an
example Moomba configuration with v0.1.1 of the Obsidian source code.
In order to focus on information that may be available in an exploration
context (i.e. publicly available geophysical surveys without contact points),
we omit the thermal sensor readings, relying on a joint inversion of
gravity, magnetic, and magnetotelluric data.

We run Obsidian's parallel-tempered sampler using 4 {simultaneous}
temperature ladders or ``stacks'' of chains, each with 8 temperatures,
as a baseline configuration.  {The posterior is formally defined in terms
of samples over the world parameters, so when quantifying predictions for
particular regions of the world and their uncertainty (such as entropy),
the parameter samples are each used to create a voxelised realization of
the 3-D world, and the average observable calculated over these voxelised
samples.}  A quantitative summary of our results is shown
in Table~\ref{tbl:metrics}, including, for each run:
\begin{itemize}
\item the shortest ($\tau_\mathrm{min}$), median ($\tau_\mathrm{med}$),
      and longest ($\tau_\mathrm{max}$) autocorrelation time measured
      for individual model parameters;
\item the standard deviations
      $\sigma_\mathrm{grav}$ and $\sigma_\mathrm{mag}$, of the gravity and
      magnetic anomaly sensor data from the posterior mean forward model
      prediction, in physical units;
\item the mean information entropy $\bar{S}$
      \citep{wellmann2012uncertainties} of the posterior probability density
      for granite, averaged over the volume beneath 3.5 km, in bits
      (i.e. presence or absence of granite; an entropy of 0~bits means
       total certainty, while 1~bit of entropy indicates total uncertainty);
\item the CPU time spent per worst-case autocorrelation time, as a measure
      of computational efficiency.
\end{itemize}

% ----------------------------------------------------------------------------

\begin{table*}[t]
\caption{Performance metrics for each run, including:
         best-case, median, and worst-case autocorrelation times
         for model parameters; standard deviations of residuals
         from the data for each sensor; volume-average information
         entropy; number of chain iterates; and CPU-hours per
         autocorrelation time.}
\begin{tabular}{lrrrrrrrrl}
\tophline
Run & $\tau_{i,\mathrm{min}}$ & $\tau_\mathrm{med}$ & $\tau_{i,\mathrm{max}}$
    & $\sigma_\mathrm{grav}$ & $\sigma_\mathrm{mag}$
    & $\bar{S}$ & $N$ & CPU (h) & Comments \\
    & ($/1000$) & ($/1000$) & ($/1000$)
    & (mgal) & (nT) & (bits) & & $/\tau_\mathrm{max}$ \\       % iterations
\middlehline % ----------------------------------------------- % to Rhat<1.1
A   &  4.3 &  16.4 &  67.8 & 0.4 & 19.2 & 0.79 & 764.5k        % ~340k
    & 10.8 & baseline iGRW \\
A1  &  4.7 &  10.7 &  42.8 & 0.4 & 18.5 & 0.68 & 1566.5k
    &  8.1 & \ldots with $N_\beta = 12$ \\
B   &  2.1 &   4.0 &  28.4 & 0.5 & 18.8 & 0.66 & 628.8k        % ~80k
    &  5.5 & baseline pCN \\
B1  &  2.4 &   4.4 &  24.3 & 0.5 & 20.5 & 0.62 & 1166.5k
    &  6.2 & \ldots with $N_\beta = 12$ \\
C   &  1.9 &  17.4 & $>143.2$ & 0.5 & 20.9 & 0.57 & 872.6k     % >600k
    & $>19.7$ & baseline aGRW \\
C1  &  2.7 &  14.1 & 310.6 & 0.4 & 17.1 & 0.61 & 2190.2k
    & 53.8 & \ldots with $N_\beta = 12$ \\
D   &  2.3 &   7.2 &  54.9 & 0.8 &  5.7 & 0.47 & 586.6k
    & 11.5 & Cauchy likelihood \\
% D1 &  4.6 &  14.0 & 244.0 & 0.9 &  6.0 & 0.46 & 1907.1k
%   & 59.6 & \ldots with $N_\beta = 12$ \\
E   &  3.0 &   8.0 & $>172.1$ & 0.7 & 6.4 & 0.51 & 669.2k
    & $>29.0$ & 5~km margin \\
F1  & 12.0 & 101.9 & $>505.6$ & 0.5 & 4.6 & 0.43 & 2386.4k
    & $>229.5$ & smoothed data, $N_\beta = 12$ \\
F4  & 13.6 &  42.1 & 170.3 & 0.6 &  7.8 & 0.54 & 3823.7k
    & 39.9 & \ldots subsampled to 100 pts/sensor \\
J   &  1.6 &  26.3 & 115.4 & 0.8 &  7.0 & 0.61 & 1172.6k
    & 11.0 & loosen rock property priors \\
J2  &  2.1 &   7.9 &  53.8 & 1.1 &  9.4 &      & 497.7k
    & 14.4 & \ldots using 1 top layer only \\
K   &  4.2 &  19.8 &  64.7 & 0.5 &  9.9 & 0.90 & 708.8k
    &  9.9 & loosen control point priors \\
K2  &  3.7 &   7.7 &  24.7 & 0.5 &  8.4 &      & 479.1k
    &  7.4 & \ldots using 1 top layer only \\
\bottomhline
\end{tabular}
% \belowtable{} % Table Footnotes
\label{tbl:metrics}
\end{table*}

% ----------------------------------------------------------------------------

\subsection{Choice of within-chain proposal}
\label{subsec:proposals}

The initial work of \citet{mccalman2014} and \citet{beardsmore2016}
used an isotropic Gaussian random walk (iGRW) proposal within each chain,
\begin{equation}
\vec{\theta}' = \vec{\theta}_{n} + \eta \vec{u},
    \hspace{0.5in} \vec{u} \sim N(\vec{0}, \vec{I}),
\end{equation}
where $\eta$ is a (possibly adaptive) step size parameter.
Each dimension of a sampled parameter vector is ``whitened'' by dividing it
by a scale factor corresponding to the allowed full range of the variable
(of order a few times the prior width; {this also accounts for
differences in physical units between parameters}).
This should at least provide a scale for the marginal distribution of each
parameter, but does not account for potential correlations between parameters.
The covariance matrix of the iGRW proposal is a multiple of the identity
matrix, so that on average, steps of identical extent are taken along every
direction in parameter space.  When tuning the proposal, the adaptive scheme
tunes only an overall step size applying to all dimensions at once.

The iGRW proposal is the simplest proposal available, but as noted above, it
loses efficiency in high-dimensional parameter spaces, and it is unable to
adapt if the posterior is highly anisotropic {--- for example,
if parameters are scaled inappropriately or are highly correlated}.
The overall step size will
adapt to the proposal width along the narrowest dimension, and the random walk
will slowly diffuse along the other dimensions; the time it takes to traverse
the entire posterior distribution should scale roughly as the square of the
condition number of the Fisher matrix.

If the global shape of the posterior is not known, it can be determined using
an adaptive/anisotropic Gaussian random walk \citep{haario2001adaptive}.
The covariance of the aGRW proposal is calculated in terms of the sample
covariance of the chain history $\{ \vec{\theta}^{[j]} \}$:
\begin{equation}
\vec{\theta}' = \vec{\theta}_{n} + \eta \vec{u},
    \hspace{0.5in} \vec{u} \sim N(\vec{0}, \vec{\Sigma}_n),
\end{equation}
in which
\begin{equation}
\vec{\Sigma}_n = \frac{n}{n+a} \, \mathrm{cov}
                                  \left\{ \vec{\theta}^{[j]} \right\}
         + \frac{a}{n+a} \, \vec{I},
\end{equation}
where $a$ is a timescale for adaptation (measured in samples).
As the length $n$ of the chain increases, the proposal will smoothly
transition from an isotropic random walk to an anisotropic random walk
with a covariance structure that reflects the chain history.

A third proposal, addressing high-dimensional parameter spaces, is the
\emph{preconditioned Crank-Nicholson} (pCN) proposal \citep{cotter2013pcn}:
\begin{equation}
\vec{\theta}_{n+1} = \sqrt{1-\eta^2} \vec{\theta}_n + \eta \vec{u},
    \hspace{0.5in} \vec{u} \sim P(\vec{\theta})
\end{equation}
with $0 < \eta < 1$ and $P(\vec{\theta})$ a multivariate Gaussian prior.
For $\eta \ll 1$, the proposal resembles a GRW proposal with small step
size, while for $\eta \sim 1$ the proposal becomes a draw from the prior.
This proposal results in a sampling efficiency that is independent of the
dimensionality of $\vec{\theta}$;
in fact, it was developed by \citet{cotter2013pcn}
to sample infinite-dimensional function spaces, arising in inversion problems
using differential equations as forward models, where the prior is specified
in the eigenbasis for the forward model operator.  In our case, the prior
incorporates the correlation between neighboring control points in the
Gaussian process layer boundaries, so we might expect that a proposal that
respects this structure would improve sampling.

Our first three runs (A, B, C) use the iGRW, pCN, and aGRW (with $a = 10$)
proposals respectively.  All three algorithms give roughly similar results
on the baseline dataset.  The autocorrelation time for this problem remains
very long, of the order of $10^4$ samples.  This means that $\sim 10^6$
samples are required to achieve reasonable statistical power.

There are nevertheless differences in efficiency among the samplers.
The pCN proposal has not only the lowest median autocorrelation,
but the lowest worst-case autocorrelation across dimensions.
The aGRW proposal has the largest spread in autocorrelation times across
dimensions, with its median performance comparable to iGRW and its worst-case
performance at least three times worse
(it had still failed to converge after over 1000~CPU-hours).
Repeat trials running for twice as many samples with 12 chains per stack
instead of 8 (Runs A1, B1, C1) produced similar results, although we were then
able to reliably measure the worst-case autocorrelation time for aGRW.
For all samplers, but most noticeably aGRW, the step size can
take a long time to adapt.  Large differences are sometimes seen in the
adapted step sizes between chains at {similar temperatures} in different
{stacks}, and do not always increase monotonically with temperature.

The differences are shown in Fig.~\ref{fig:traceplot},
showing the zero-temperature chains from the four stacks in each run
{sampling the marginal distribution of}
the rock density for layer 3, a bimodal parameter.
The iGRW chains converge slowly, and though they manage to travel between
modes with the help of parallel-tempered swap proposals, the relative weights
of the two modes are not fully converged and vary between re-runs
at a fixed length.  Each aGRW chain has a relatively narrow variance and none
successfully crosses over to the {high-density} mode despite
parallel-tempered swaps.  Only the pCN chains converge ``quickly''
(after about 70k samples) and are able to explore the full width
of the distribution.

\begin{figure}[t]
\includegraphics[width=8.3cm]{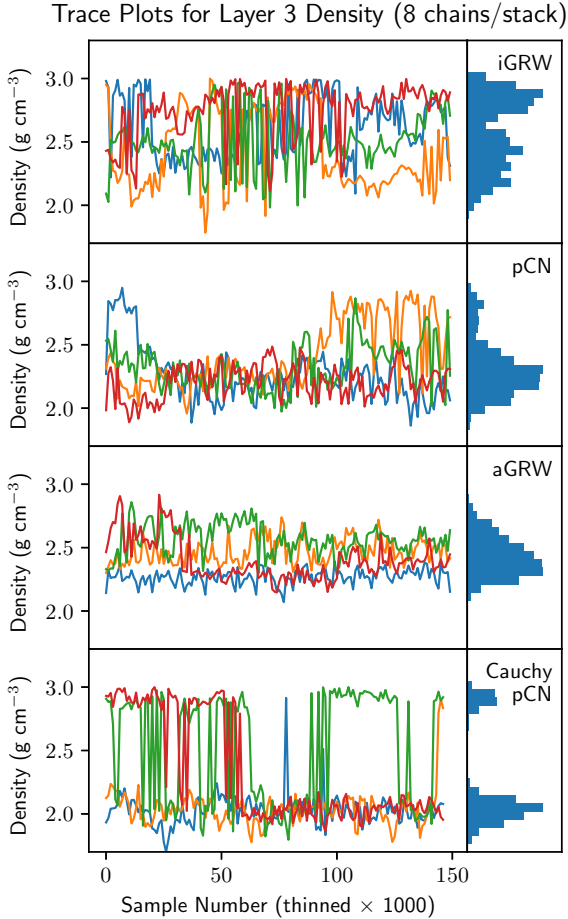}
\caption{Trace plots (left) and marginal densities (right) for layer 3
         rock density as explored by iGRW, pCN, and aGRW proposals,
         and by a pCN proposal under a Cauchy likelihood (top to bottom).
         The four colors represent the four different chains.}
\label{fig:traceplot}
\end{figure}

These behaviors suggest that the local shape of the posterior varies across
parameter space, so that proposals that depend on a global fixed scaling
across all dimensions are unlikely to perform well.  The clearly superior
performance of pCN for this problem is nevertheless intriguing, since for
sufficiently small step size near $\beta = 1$, the proposal reduces to GRW.

Figure~\ref{fig:traceplot} shows that iGRW and aGRW have more trouble
traveling between different posterior modes than pCN.  This is true despite
the fact that all three proposals are embedded within a PTMCMC scheme with
a relatively simple multivariate Gaussian prior, to which aGRW should be able
to adapt readily.
We believe pCN will prove to be a good baseline proposal for tempered sampling
of high-dimensional problems because of its prior-preserving properties,
which ensure peak performance when constraints from the data are weak.
As the chain temperature increases, the tempered posterior density approaches
the prior, so that pCN proposals with properly adapted step size will smoothly
approach independent draws from the prior with an acceptance probability of 1.
The result is that when used as the
within-chain proposal in a high-dimensional PTMCMC algorithm, pCN proposals
will result in near-optimal behavior for the highest-temperature chain,
and should explore multiple modes much more easily than GRW proposals.

This behavior stands in contrast to GRW proposals, for which the acceptance
fraction given any particular tuning will approach zero as the dimension
increases.  In fact, aGRW's attempt to adapt globally to proposals with local
structure may mean mid-temperature chains become trapped in low-probability
areas and break the diffusion of information down to lower temperatures
from the prior.  A more detailed study of the behavior of these proposals
within tempered sampling schemes would be an interesting topic for future
research.

% ----------------------------------------------------------------------------

\subsection{Variations in likelihood / noise prior}
\label{subsec:likelihood}

% Plots:  contours, residuals, voxel slices for runs {B, D}

In the fiducial Moomba configuration, the priors on the unknown variance of
the Gaussian likelihood for each sensor are relatively informative.
For example, the choice $\alpha = 5$, $\beta = 0.5$ --- used for the gravity
and magnetotelluric sensors --- corresponds to noise with standard deviation
$21\% < \sigma < 52\%$ with 95\% probability, and a median of $\sigma = 32\%$
(that is, as a percentage of the sample standard deviation of the data in its
original units).  The resulting $t$-distribution for each data point has
$\nu = 2\alpha = 10$ degrees of freedom, so that the non-Gaussian tails
resulting from an imprecisely known noise variance are strongly suppressed.
Thus the likelihood is close to being Gaussian with fixed variance 0.32.
The magnetic sensor, on the other hand, uses $\alpha = 1.25$, $\beta = 1$,
a much more permissive prior.

If specific informative prior knowledge about observational errors exists,
using such a prior, or even fixing the noise level outright, makes sense.
In cases where the amplitude of the noise term is not well-constrained,
using a broader prior on the noise term may be preferable.  When more than
one sensor with unknown noise variance is used, identical broad priors allow
the data to constrain the relative influence of each sensor on the final
results.  The trade-off is that a more permissive prior on the noise variance
could mask structured residuals due to model inadequacy or non-Gaussian
outliers in the true noise distribution.

The idea that such broad assumptions could deliver competitive results
arises from the incorporation of Occam's razor into Bayesian reasoning,
as demonstrated in \citet{sambridge2012}.  For example, the log likelihood
corresponding to independent Gaussian noise is
\begin{equation}
\log \mathcal{L} = -\frac{1}{2} \sum_{j=1}^{N_d}
    \left[ \frac{(f_{sj}(\vec{\theta}) - \mathcal{D}_{sj})^2}{\sigma^2}
           + \log \, 2\pi\sigma^2 \right].
\end{equation}
Ordinary least-squares fitting maximizes the left-hand term inside the sum,
and the right-hand term is a constant that can be ignored if the observational
uncertainty $\sigma$ is known.  This clearly penalizes worlds $\vec{\theta}$
resulting in large residuals.  Suppose that $\sigma$ is unknown, however,
and is allowed to vary alongside $\vec{\theta}$:  the left-hand term penalizes
small (overly confident) values of $\sigma$, while the right-hand term
penalizes large values of $\sigma$ corresponding to an assumption that the
data are entirely explained by observational noise.

Typical residuals from the fiducial inferences correspond to about 10\%
of the dataset's full range, so we next perform a run in which
the noise prior is set to $\alpha = 0.5, \beta = 0.05$ for all samples.
The corresponding likelihood (with the noise variance prior integrated out)
becomes a Cauchy (or $t_1$) distribution,
with thick tails that allow substantial outliers from the core.
This choice of $\alpha$ and $\beta$ thus also allows us to make contact with
prior work where Cauchy distributions have been used
\citep{silva1989,gempy}:
a Gaussian likelihood with unknown, $\mathrm{IG}(0.5,\beta_s)$-distributed
variance is mathematically equivalent to a Cauchy likelihood with known scale
$2\beta_s$.  The two choices are conceptually different, since in the Gaussian
case outliers appear when the wrong variance scale is applied, whereas in the
Cauchy case the scale is assumed known and the data have an intrinsically
heavy-tailed distribution.

Under this new likelihood the residuals from the gravity observations increase
(by about a factor of 1.5--2), while the residuals from the magnetic sensors
decrease (by a factor of 3--4).  This rebalancing of residuals among the
sensors with an uninformative prior can be used to inform subsequent rounds
of modelling more readily.

The inference also changes:  in run~D, a granite bridge runs from the main
outcrop to the eastern edge of the modelled volume, with the presence of
granite in the northwest corner being less certain.  Agreement with run~B and
with the \citet{beardsmore2016} map is still good along the eastern edge.
The posterior entropy also decreases substantially, due to increase in the
probability of granite structures at greater depths (beneath 3.5~km).

The weight given to the gravity sensor is thus
an important factor determining the behavior of the inversion throughout half
the modeled volume.  With weakened gravity constraints, the two modes for the
inferred rock density in layer 3 separate widely
(see Fig.~\ref{fig:traceplot}), though the algorithm is still able to move
between the modes occasionally.  The marginal distributions of the other
rock properties do not change substantially, and remain unimodal.

The comparison map for the inversion of \citet{beardsmore2016} comes from the
deterministic inversion of \citet{meixner2009}, which uses gravity as the main
surface sensor but relies heavily on seismic data, with reflection horizons
used to constrain the depth to basement, and measurements of wave velocities
(which correlate with density) from a $P$-wave refraction survey to constrain
density at depth.  While \citet{meixner2009} mention constraints on rock
densities, no mention is made of the level of agreement with the gravity data.

Without {more information} --- a seismic sensor in our inversion,
priors based on the specific seismic interpretations of \citet{meixner2009},
{or specific knowledge about the noise level in the gravity dataset
that would justify an informative prior} --- it is hard to say how concerned
we should be about the differences between the deterministic inversion
and our probabilistic version.  
The comparison certainly highlights the potential importance of seismic data,
both as a constraint on rock properties at depth and on the geometry of
geological structures.

{Indeed, one potential weakness of this approach to balancing sensors is
model inadequacy:  the residuals from the inference may systematic residuals
from unresolved structure in the model, in addition to sensor noise.
The presence of such residuals
is a model selection question that in a traditional inversion context would
be resolved by comparing residuals to the assumed noise level, but this
depends strongly upon informative prior knowledge of the sensing process for
\emph{all sensors} used in the inversion.}
The remaining experiments will use the Cauchy
likelihood unless otherwise specified.

\begin{figure*}[t]
\includegraphics[width=12cm]{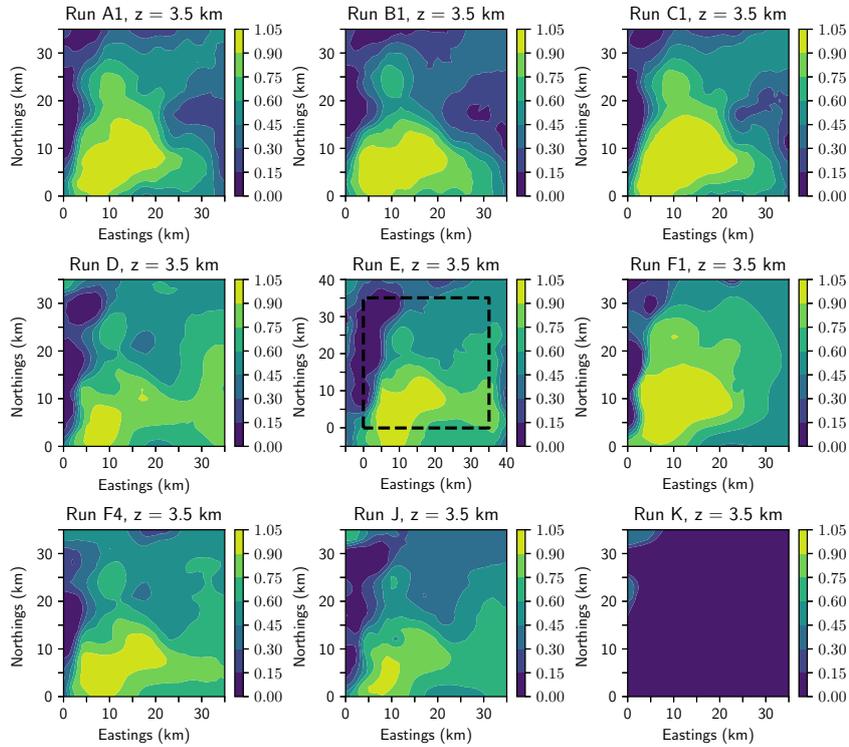}
\caption{Slices through the voxelised posterior probability of occupancy
         by granite for each run at a depth of 3.5~km.}
\label{fig:moomba-slices}
\end{figure*}

\begin{figure*}[t]
\includegraphics[width=12cm]{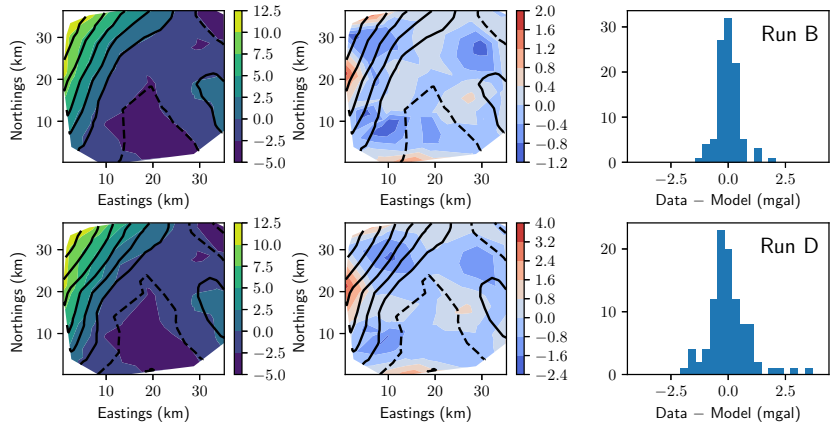}
\caption{Gravity anomaly at the surface ($z = 0$).
         In a contour plot (left):  filled contours = observations,
         black lines = mean posterior forward model prediction.
         Residuals of observations from the mean posterior forward model
         are also shown as a contour map (middle) and histogram (right).}
\label{fig:contours-BD-grav}
\end{figure*}

\begin{figure*}[t]
\includegraphics[width=12cm]{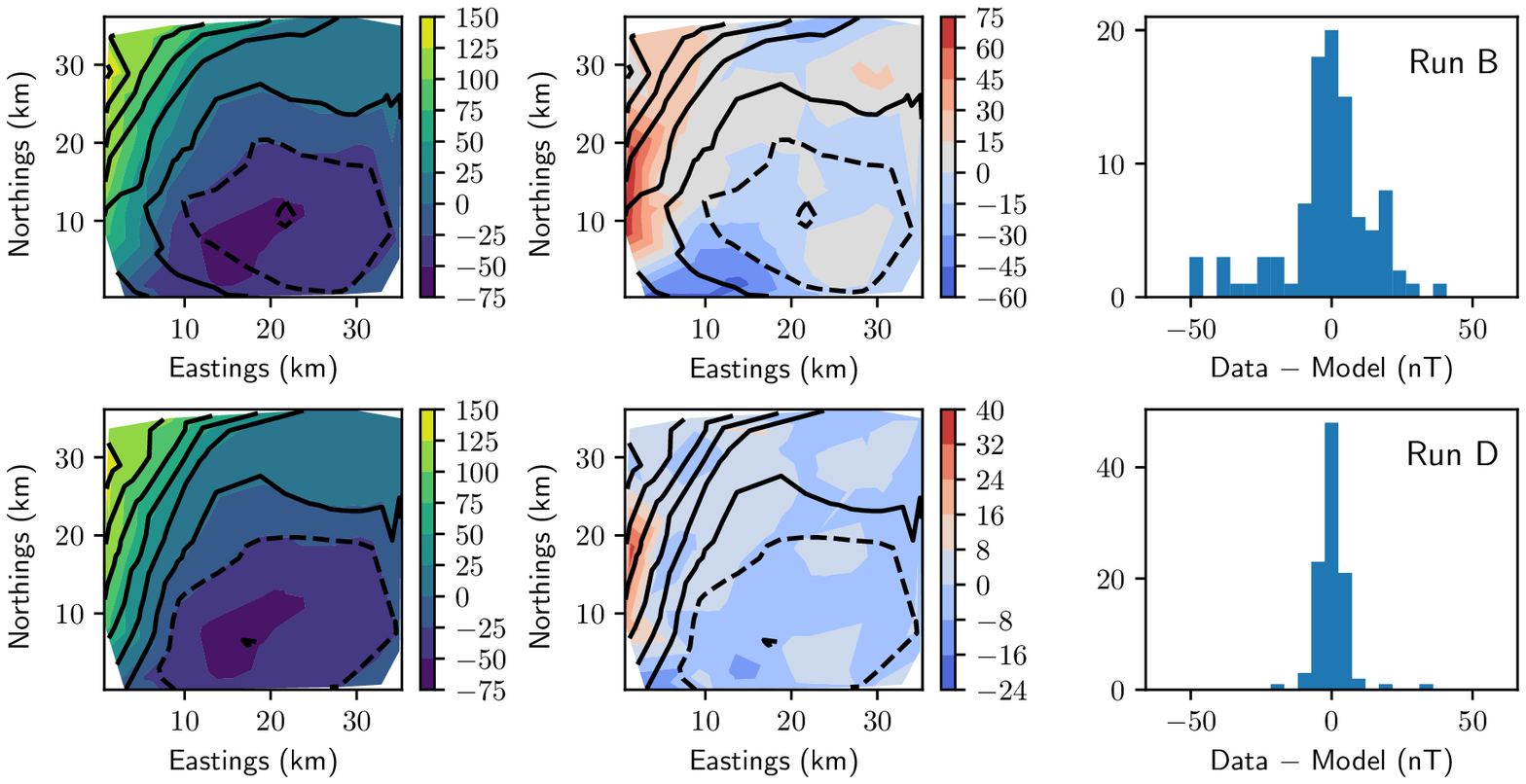}
\caption{Magnetic anomaly at the surface ($z = 0$)
         in the same format as Fig.~\ref{fig:contours-BD-grav}.}
\label{fig:contours-BD-mag}
\end{figure*}

% ----------------------------------------------------------------------------

\subsection{Boundary conditions}
\label{subsec:boundary}

% Plots:  none needed

The boundary conditions Obsidian imposes on {world voxelisations}
assume that rock properties rendered at a boundary edge
(north/south, east/west) extend indefinitely off the edges, e.g.
\begin{eqnarray*}
\rho_{is}(x < {x}_\mathrm{min}) = \rho_{is}({x}_\mathrm{min}),\\
\rho_{is}(x > {x}_\mathrm{max}) = \rho_{is}({x}_\mathrm{max}).
\end{eqnarray*}
This may not be a good approximation when rock properties show strong
gradients near the boundary.
The residual plots shown in Fig.~\ref{fig:contours-BD-grav}
and \ref{fig:contours-BD-mag} show persistently high
residuals along the western edge of the {world}, where such gradients
appear in both the gravity anomaly and the magnetic anomaly.

For geophysical sensors with localized response, one way to mitigate this is
to {include in the world representation}
a larger area than the sensor data cover, incorporating a margin with
width comparable to the scale of boundary artifacts, in order to let the model
respond to edge effects for sensors with a finite area of response.  In run~E,
we add a boundary zone of width 5~km around the margins of the {world},
while increasing the number of control points in the granite intrusion layer
boundary from 49 ($7 \times 7$ grid) to 64 ($8 \times 8$ grid).
Neither the model residuals nor the inferred rock geometry substantially
differs from the previous run, suggesting that the remaining outliers are
actual outliers and not primarily due to mismatched boundary conditions.
The autocorrelation time, however, increases substantially due both to the
increase in the problem dimension and the fact that the new world parameters
are relatively unconstrained, hence poorly scaled relative to the others.

% ----------------------------------------------------------------------------

\subsection{Smoothed/resampled sensor data}
\label{subsec:smoothing}

% Plots:  contours + residuals for run F, F1, F2, or whatever other ones

The Obsidian likelihood assumes that the observational noise fluctuations
in the sensor data are statistically independent.  This is also the implicit
assumption behind least-squares fitting, but it may not be true if the data
have been interpolated, resampled, or otherwise modified from original point
observations.  For example, gravity anomaly and magnetic anomaly
measurements are usually taken at ground level along access trails to a site,
or along spaced flight lines in the case of aeromagnetics.
In online data releases, the original measurements may then be interpolated
or resampled onto a grid, changing the number and spacing of points and
introducing correlations on spatial scales comparable to the scale of
the smoothing kernel.
{This resampling of observations onto a regular grid may be useful
for traditional inversions using Fourier transform techniques.
However, if used uncritically in a Bayesian inversion context,
correlations in residuals from the model may then arise from the resampling
process} rather than from model {misfit},
resulting in stronger penalties in the likelihood for what would otherwise
be plausible worlds, and muddying questions around model inadequacy.

To simulate these effects, we interpolated the original, irregularly sampled
gravity and magnetic anomaly data onto regular grids with 1.5~km spacing,
resulting in 552 samples for each dataset
(from the original observations, as there would be no way after
the regridding to tell how many independent observations there had been).
For each interpolation we used the maximum \emph{a posteriori} fit
of a Gaussian process regression with a square exponential kernel;
the best-fit correlation length was 8~km (7~km) with residuals of about
4\% (2\%) of standard deviation for the gravity (magnetic) anomaly data.
To the extent that these potential-field sensors represent moving averages 
of the underlying rock properties on some length scale, these results show
why a $7 \times 7$ grid of control points (spacing 4.4~km) should provide
adequate resolution for reconstruction of bulk geology at
{3.5~km target} depth.
Although different interpolation schemes, such as linear, loess, or spline
interpolation, may be applied for resampling and smoothing
in contemporary online surveys, Gaussian process regression can be easily
applied to irregularly sampled data and so forms a reasonable test case here.

The residuals from the original data are well below the typical residuals
of model inadequacy based on other runs.  However, the median and worst-case
autocorrelation times are over 10 times longer than for the unsmoothed data
of Run~D, and indeed the run failed to converge before the adaptation
decreased to negligible influence.  The non-converged 
Run~F posterior looks different, with a much
higher probability of granite at 3.5~km depth than the other inversions,
together with lower volume entropy in the inversion overall.
The bridge characterizing Runs D and E (with the same likelihood) vanishes.
Despite the superficial improvement in residuals, we might well view
the results with suspicion, since information has been taken out of the data.

Variations on this standard case make little difference.  Running with up to
16 parallel-tempered chains, thereby doubling the computation time, produces
similar results.  Making the noise prior more restrictive, in case the high
level of correlation is due to enabling exploration of too large a space
(for example $\alpha = 10$, $\beta = 0.3$, a roughly Gaussian likelihood
with fixed variance of 3\%) results in higher MAP probability
but no improvement in sampling.  Adding random Gaussian noise of 3\% to
the smoothed data, to match the fluctuations around the original data and
satisfy the assumptions of the noise prior, also has no obvious effect.
Fitting a random subsample of 100 points each from the smoothed gravity and
magnetic anomaly datasets, producing smoothed datasets with the same size
and approximate distribution as the original data (Run~F4),
reduces the autocorrelation time by more than half, but it remains six times
as long as Run~D.

The reprocessing of the data has had several effects:  First, it has increased
the number of (assumed independent) data points to fit, tightening constraints
where no new information has been added and making the posterior more
difficult to explore.  Second, even after subsampling the dataset to match
the original number of points, some correlations still remain; the
probability that neighboring points will deviate from the fit in the same
direction is larger than for the unsmoothed data.  Thus, while spatially
coherent residuals might ordinarily point to inadequacy in the
world {parametrization},
the results are unclear if the data have been smoothed.  A kernel smoothing
radius of 7~km results in at most (35~km/7~km)$^2$ = 25 independent spatial
regions, so this is the effective sample size of our smoothed data.  The loss
of information means that not only is the answer more uncertain (even biased),
but the algorithm mistakenly reports a \emph{less} uncertain answer through
the smaller posterior variance.

This cautionary tale shows that for best results, the input data should not
be smoothed, or should at least be subsampled to reduce correlation between
points (if the correlation scale is known).  Improved results could also be
obtained by using a multivariate Gaussian likelihood with correlations on
the appropriate spatial scale (that is, a Gaussian process likelihood).
A Gaussian process likelihood, however, complicates matters by introducing a
length scale hyperparameter into the sampling, and by risking confusion
between spatially coherent model errors and correlated observational noise.

% ----------------------------------------------------------------------------

\subsection{Looser priors on rock properties and layer depths}
\label{subsec:priors}

In cases where samples of rock for a given layer are few or unavailable,
the empirical covariance used to build the prior on rock properties may be
highly uncertain or undefined.  In these cases, the user may have to resort to
a broad prior on rock properties.  The limiting case is when no petrophysical
data are available at all.  Similarly, definitive data on layer depths may
become unavailable in the absence of drill cores, or at least seismic data,
so that a broad prior on control point depths may also become necessary.

We re-run the main Moomba analysis using two new priors.  The first (run~J)
simulates the absence of petrophysical measurements.  The layer depth priors
are the same as the fiducial setup, but the rock property prior for each
layer is now replaced by an independent Gaussian prior on each rock property,
with the same mean as in previous runs but a large width common to all layers:
\begin{equation}
\rho_{is} \sim \mathcal{N}(\mu_{\rho_{is}}, \sigma_{\rho_{is}}).
\end{equation}
The standard deviations are 0.2~g~cm$^{-3}$ (density),
0.5 (log magnetic susceptibility), and 0.7 (log resistivity in $\Omega$~m).

The Run~J voxelisation shows reasonable correspondence with the baseline
run~D, though with larger uncertainty, particularly in the northwest corner.
In the absence of petrophysical samples, but taking advantage of priors on
overlying structure from seismic interpretations, a preliminary segmentation
of granite from basement can thus still be obtained using broad
priors on rock properties.  Although the algorithm cannot reliably infer the
bulk rock properties in the layers, the global prior on structure is enough
for it to pick out the shapes of intrusions by looking for \emph{changes} in
bulk properties between layers.

The second run (run~K) removes structural prior information instead of
petrophysical prior information.  The priors on rock properties are as in
the fiducial setup, but the control point prior for each layer is replaced
by a multivariate Gaussian with the same anisotropic Gaussian covariance,
\begin{equation}
\vec{\Sigma}_{\vec{\alpha}} = \sigma_\alpha \left[
    % \begin{array}{cccc}
    \begin{matrix}
        1.0 & 0.5 & \ldots & 0.5 \\
        0.5 & 1.0 & \ldots & 0.5 \\
        \vdots & \vdots & \ddots & \vdots \\
        0.5 & 0.5 & \ldots & 1.0
    \end{matrix}
    % \end{array}
\right]
\end{equation}
with $\sigma_\alpha = 3$~km.

Run~K yields no reliable information about the location
of granite {at 3.5~km depth}.  This seems to be due solely to the
uncertain thickness of layers of sedimentary rock that are constrained to be
nearly uniform horizontal slabs in Run~J, {corresponding to a known
insensitivity to depth among potential-field sensors.}
When relaxed, these strong priors cause a crisis of identifiability for the
resulting models.  {Further variations on Runs~J and K} show that
replacing these multiple thin layers with a single uniform slab of
$\sim 3$~km depth {(Runs J2 and K2)} does not aid either convergence
or accuracy, as long as more than one layer boundary is allowed to have
large-scale structure.

As mentioned above and in \citet{beardsmore2016}, the strong priors on layer
boundaries and locations were originally derived from seismic sensor data.
Such data will not always be available, but seem to be critical to constrain
the geometry of existing layers to achieve a plausible inversion at depth.

% Extensions:  Gaussian mixtures on rock properties, hierarchical priors
% for rock properties (especially re: potentially unknown number of layers)

% ============================================================================

\section{Discussion}

% Things we found out:
% -- Random-walk samplers are all pretty bad at these posteriors;
%    the limiting factor is local structure from weird priors
% -- Knowing about weaknesses in your sensors, like mass-sheet degeneracy
%    for potential-field sensors, is critical to setting up the problem
% -- Seismic data is actually really important to get, and structural
%    measurements at surface likely just as important (see GemPy)
% -- Try to acquire *some* idea of your noise properties, and acquire
%    original un-gridded data for best results
%
% What are the main take-away points?
% -- Care must be taken in just setting up the problem, as seemingly
%    innocuous things like wanting to be robust to outliers can backfire
% -- Some challenges remain, like just sampling the damn model
% -- Likely advances from:
%    ** more faithful parametric representations of worlds, that are both
%       geologically plausible and more parsimonious than non-parametric fits
%    ** better sampling schemes based on smooth surrogate posteriors
%    ** faster surrogate forward models

The clearest lesson we can draw from the various inversions we have run is
that the posterior uncertainty can be much larger than one might expect from
point-estimate or deterministic inversions.  Our results were sensitive
to the MCMC proposal used (in that some proposals were extremely inefficient
and gave wrong results if stopped early; see Fig.~\ref{fig:traceplot});
to the assumed weighting given to different sensors;
to the way in which data might have been pre-processed;
and to the quality and quantity of informative prior information.

The changes in the posterior under different priors are not always intuitive:
unrealistically tight constraints can hamper sampling, but relaxing priors
may sometimes widen the separation between modes
(as shown in Fig.~\ref{fig:traceplot}), which also makes the posterior
difficult to sample.  Additionally, particular weaknesses in sensors,
such as the {insensitivity of potential-field sensors to the depth of
geological features or to the addition of any horizontally invariant density
distribution}, can make it impossible to distinguish using those
data between multiple plausible alternatives, adding to the irregularity and
multi-modality of the posterior.

While any single data source may be easy to understand on its own,
unexpected interactions between parameters can also arise.  Structural priors
from seismic data or geological field measurements appear to play a crucial
role in stabilizing the inversions in this paper, as seen by the collapse
of our inversion after relaxing them.

%I think the above paragraphs needs to be reorganized a bit. Maybe one issue per paragraph? one on MCMC sampling, the other on sensors - but should refer back to your experiments - research questions?

Our findings reinforce the impression  that to make Bayesian inversion
techniques useful in this context, the computational burden must be
reduced by developing efficient sampling methods.  Three complementary ways forward present themselves:
\begin{enumerate}
\item to develop MCMC proposals, or non-parametric methods to approximate
      probability distributions, that both function in (relatively)
      high-dimensional spaces and capture local structure in the posterior;
\item to develop fast approximate forward models for complex sensors
      (especially seismic) that deliver detailed information at depth,
      along with new ways of assessing and reducing model inadequacy;
\item to develop richer world parametrizations of 3-D geological models
      that faithfully represent real-world structure in as few dimensions
      as possible.
\end{enumerate}
All three of the MCMC proposals studied here are variations of random walks,
which explore parameter space by diffusion and do not easily handle posteriors
with detailed local covariance structure such as the ones we find here.
Proposals that can sense and adjust to local structure from the present state
require, almost by definition, knowledge of gradients \citep{neal2011mcmc}
%https://arxiv.org/pdf/1011.6217.pdf
%https://pdfs.semanticscholar.org/103f/9bae3812b0d15c721152d99def1544312996.pdf  - perhaps this could be cited for future easy amendments to RW proposals, if you not done already
or higher-order curvature tensors \citep{girolami2011riemann},
which in turn require gradients of both the prior and the likelihood
(in particular, of forward models).

Forming gradients of forward models by finite differences is likely to be as
prohibitively expensive as not having gradients; furthermore, practitioners
may not have the luxury of rewriting their forward model code to return
gradients.  This is one goal of writing fast emulations of forward models,
particularly emulations for which derivatives can be calculated analytically
{\citep[see for example][]{fichtner2006adjoint1,fichtner2006adjoint2}}.
Smooth universal approximators, such as artificial neural networks, are one
possibility; Gaussian process latent variable models \citep{titsias2010gplvm}
or Gaussian process regression networks \citep{wilson2011gprn}
are others, which would also enable nonlinear dimensionality reduction for
difficult forward models or posteriors.
reduction.  Algorithms that alternate between
fast/approximate forward models for local exploration, on the one hand,
and expensive/precise forward models for evaluation of the objective function,
on the other, have proved useful in engineering design problems
\citet{jin2011,sobester2014}.
These approximate emulators give rise to model inadequacy terms in the
likelihood, which can be {explicitly} addressed; for example,
\citet{kopke2018} present a
geophysics inversion framework in which the inference scheme learns a model
inadequacy term as sampling proceeds, {showing proof of principle
on a crosshole georadar tomography inversion}.
A related, complementary route is to produce analytically differentiable
approximations to the posterior, built as the chain explores the space
\citep{strathmann2015kmc,lan2016}.

Another source of overall model inadequacy comes from the world
parametrization  which can be viewed as part of the prior.
Obsidian is tuned to match sedimentary basins; its world
{parametrization}
is too simple to represent {more complex structures,}
particularly those with {abrupt variations caused, for example,
by fault displacement.}
The GemPy package developed by \citet{gempy} makes an excellent start on a
more general-purpose package.  GemPy is also specifically written to take
advantage of autodifferentiation, providing ready gradient information for
the prior.

% [RC: need to think of what more can be done in future work.
%  Perhaps, use of the approach for a different study area.
%  Further enhancements to PT-MCMC - in terms of incorporation of
%  evolutionary optimisation in PT  to form better non-gradient based
%  proposals - that considers a population of recent samples.]}

% ============================================================================

\conclusions

We have performed a suite of 3-D Bayesian geophysical inversions for the
presence of granite at depth in the Moomba gas field of the Cooper basin,
altering aspects of the problem setup to determine their effects on the
efficiency and accuracy of MCMC sampling.  Our main findings are as follows:
\begin{itemize}
\item Parametrized worlds have much lower dimensionality than non-parametric
      worlds, and the parameters also offer a more interpretable description
      of the world --- for example, boundaries between geological units are
      explicitly represented.  However, the resulting posterior has complex
      local covariance structure in parameter space, even for linear sensors.
\item Although isotropic random walk proposals explore such posteriors
      inefficiently, poorly adapted anisotropic random walks are even less
      efficient.  A modified high-dimensional random walk such as pCN
      outperforms these proposals, and the prior-preserving properties of
      pCN make it especially attractive for use in tempered sampling.
      However, proposals using gradients from autodifferentiation are
      probably needed to make further progress in this area.
\item The shape of the posterior and number of modes can also depend in
      complex ways upon the prior, making tempered proposals essential.
\item In cases where the relative observational noise levels in the data
      are not well-constrained, using identical, uninformative priors on
      the noise level for each sensor allows the inversion algorithm to
      rebalance information among sensors for a better fit.
\item Smoothing or resampling sensor data leads to loss of sampling
      efficiency as well as a biased, unrealistically certain posterior.
      In these cases, subsampling the data to reduce correlations can
      aid sampling.  The introduction of correlations into the likelihood
      may also improve the accuracy of the posterior, although sampling may
      still be inefficient without a properly tuned MCMC proposal.
\item Useful information about structures at depth can sometimes be obtained
      through sensor fusion even in the absence of informative priors.
      However, direct constraints on 3-D geometry from seismic
      interpretations or structural measurements
      seem to play a privileged role among priors,
      {owing to the relatively weak constraints on depth of structure
            afforded by potential field methods.}
\end{itemize}
In summary, both advanced MCMC methods and careful attention to the properties
of the data are necessary for inversions to succeed.

% ============================================================================

%% The following commands are for the statements about the availability of
%% data sets and/or software code corresponding to the manuscript.
%% It is strongly recommended to make use of these sections in case data sets
%% and/or software code have been part of your research the article is based on.

%% use this section when having only software code available
% \codeavailability{TEXT}

%% use this section when having only data sets available
% \dataavailability{TEXT}

%% use this section when having data sets and software code available
\codedataavailability{
The code, including configuration files, and datasets used in this paper
are available at {\mbox{\url{http://rscalzo.github.com/obsidian/}}}.
}

%% use this section when having geoscientific samples available
% \sampleavailability{TEXT}

% \appendix
% \section{}    %% Appendix A
% 
% \subsection{}     %% Appendix A1, A2, etc.

\noappendix       %% use this to mark the end of the appendix section

%% Regarding figures and tables in appendices, the following two options are possible depending on your general handling of figures and tables in the manuscript environment:

%% Option 1: If you sorted all figures and tables into the sections of the text, please also sort the appendix figures and appendix tables into the respective appendix sections.
%% They will be correctly named automatically.

%% Option 2: If you put all figures after the reference list, please insert appendix tables and figures after the normal tables and figures.
%% To rename them correctly to A1, A2, etc., please add the following commands in front of them:

% \appendixfigures  %% needs to be added in front of appendix figures

% \appendixtables   %% needs to be added in front of appendix tables

%% Please add \clearpage between each table and/or figure. Further guidelines on figures and tables can be found below.

%% optional section
\authorcontribution{ % see https://casrai.org/credit/
The study was conceptualized by SC, who with MG provided funding and
    resources.
RS was responsible for project administration and designed the methodology
    under supervision from SC and GA.
RS and DK carried out development of the Obsidian code resulting in v0.1.2,
    carried out the main investigation and formal analysis,
    and validated and visualized the results.
RS wrote the original draft text, of which all co-authors provided review
    and critical evaluation.
}

%% this section is mandatory even if you declare that no competing interests
%% are present
\competinginterests{The authors declare that they have no conflict of interest.}

% \disclaimer{TEXT} %% optional section

\begin{acknowledgements}
This work is part of the Lloyd’s Register Foundation -- Alan Turing
Institute Programme for Data-Centric Engineering.  RS thanks Lachlan McCalman,
Simon O'Callaghan, and Alistair Reid for useful discussions about the
development of Obsidian up to v0.1.1.
\end{acknowledgements}

%% REFERENCES

%% The reference list is compiled as follows:

% \begin{thebibliography}{}

% \bibitem[AUTHOR(YEAR)]{LABEL1}
% REFERENCE 1

% \bibitem[AUTHOR(YEAR)]{LABEL2}
% REFERENCE 2

% \end{thebibliography}

%% Since the Copernicus LaTeX package includes the BibTeX style file copernicus.bst,
%% authors experienced with BibTeX only have to include the following two lines:

\bibliographystyle{copernicus}
\bibliography{ref.bib}

\end{nolinenumbers}
\end{document}